\documentclass[pre,floatfix,twocolumn, 10pt]{revtex4} 
\usepackage{graphicx}
\usepackage{amssymb}
\usepackage{natbib}
\usepackage{dcolumn}
\usepackage{bm}
\usepackage{color}
\usepackage[colorlinks=true, linkcolor=blue, citecolor=blue]{hyperref}
\usepackage{subfigure}
\usepackage{graphicx,psfrag,xspace}
\usepackage{color}
\usepackage{amssymb,amsfonts,amsmath}
\usepackage{setspace}

\DeclareMathOperator\atan{atan}

\begin{document}


\author{E. A. Jagla} 
\affiliation{Centro At\'omico Bariloche, Instituto Balseiro, 
Comisi\'on Nacional de Energ\'ia At\'omica, CNEA, CONICET, UNCUYO,\\
Av.~E.~Bustillo 9500 (R8402AGP) San Carlos de Bariloche, R\'io Negro, Argentina}

\title{Volume-shear coupling in a mesoscopic model of amorphous materials}
\begin{abstract} 
We present a two-dimensional mesoscopic model of  a yield stress material that includes the possibility of local volume fluctuations coupled to shear, in such a way that the shear strength of the material decreases as the local density decreases. The model reproduces a number of effects well known in the phenomenology of this kind of materials. Particularly, we find that: the volume of the sample increases as the deformation rate increases; shear bands are no longer oriented at 45$^\circ$ with respect to the principal axis of the applied stress (as in the absence of volume-shear coupling); homogeneous deformation becomes unstable at low enough deformation rates if volume-shear coupling is strong enough. We also analyze the implications of this coupling in the context of out of equilibrium shear bands appearing for instance in metallic glasses.

\end{abstract}

\maketitle

\section{Introduction}

Dilatancy \cite{dilatancy,nedderman} is a property of granular and other kind of amorphous materials, characterized by a volume increase that is observed when the  material is forced to shear. Traditionally, dilatancy has been known to play an important role in the mechanics of soils and sands\cite{culling,dietrich,roering,jerolmack}. However, dilatancy may be considered as a particular case of 
volume-shear coupling phenomena, that are also relevant for instance in the physics of yield 
stress fluids (most of which in fact have some kind of  ``grains" as elementary constituents).\cite{berthier,nicolas,simple_yield_stress_fluids,besseling,coussot}
Also a volume change, and therefore a volume-shear coupling has been invoked as one important ingredient of the physics of materials failing by the nucleation of shear bands, particularly metallic glasses. \cite{tang,greer,ogata,zhong}

We investigate here the dilatancy effect and the volume-shear coupling in a mesoscopic model of the yielding transition. 
Previously\cite{jagla_2007,jagla1,jagla2}, this model has been applied to situations where the local density of the system was not considered to play an important role, and actually the possible change of density was not even considered. Here, we study the coupled evolution of local strain and local density, and address the influence of local density fluctuations on the shearing behavior of the system.

The paper is organized as follows. In the next section we consider a prototypical one variable, mean field model of the yielding behavior and add to it an additional variable describing density changes. We obtain in this simple model the basic properties of the dilatancy effect, particularly, the increase in system volume with the increase of strain rate in the system. Then in Section III we show how this behavior justifies that in spatially extended models the geometric orientation of plastically deformed regions, or shear bands, is rotated with respect to the case in which there is no such volumetric effect. The orientation we observe 
departs from the one predicted by the classical Mohr-Coulomb failure criterion, instead it is consistent with the orientation predicted by a recently proposed theory.
In section IV we show that volume-shear coupling can produce a flowing state in the material in which parts of the sample yield and acquire a lower density, while other parts remain essentially rigid, and with a higher density. While all previously mentioned results correspond to equilibrium and reversible situations, in Section V we investigate the important case in applications of samples prepared by some sort of annealing, which are then submitted to a shear deformation until they fail, typically by nucleating a non-equilibrium shear band, as it occurs in metallic glasses. Finally, in Section VI, we summarize and conclude.

\section{Coupling volume and shear in a one-site model}

\begin{figure}
\includegraphics[width=7cm,clip=true]{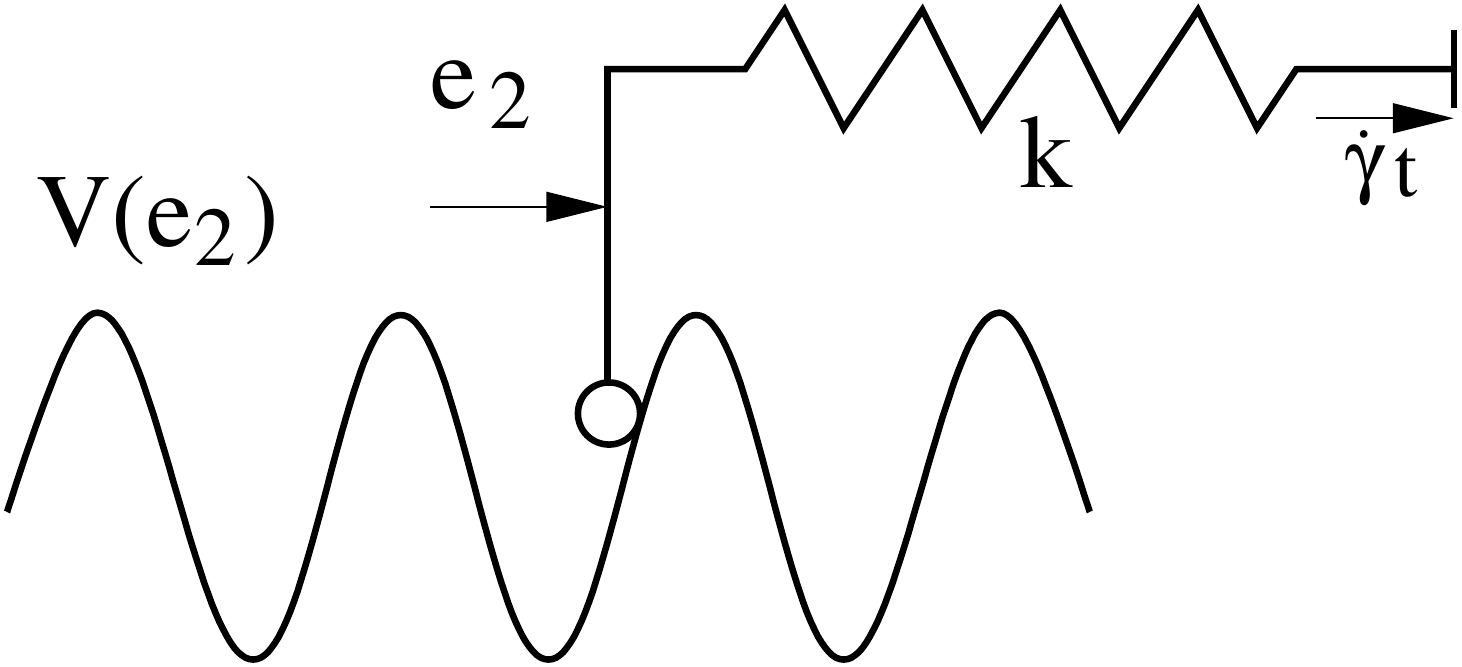}
\caption{Schematic representation of the Prandtl-Tomlinson model, mathematically described by Eq. (\ref{pt1}).}
\label{esquema_pt}
\end{figure}

We start by considering a very simple model with two degrees of freedom (representing the shear and volume state of a sample) and study the appearance of a dilatancy effect. We take as a starting point the well known Prandtl-Tomlinson model of friction for the shear degree of freedom $e_2$\cite{popov}. The model is written as a dynamical evolution equation of the form
\begin{equation}
\dot e_2= -\frac{dV(e_2)}{de_2}+(\dot\gamma t-e_2)k
\label{pt1}
\end{equation}
and it describes the situation depicted in Fig. \ref{esquema_pt}. The driving $\dot\gamma t$ pulls from the variable $e_2$ through a spring of stiffness $k$. $e_2$ is also affected by the force $f_2(e_2)\equiv-dV/de_2$. In a friction context $f_2$ represents the corrugated potential between two surfaces sliding against each other. In a context of yielding, $f_2$ represents the internal stress that a small portion of the system feels due to its amorphous nature. In any case, the $f_2(e_2)$ function is expected to have many minima. To fix ideas we will assume here that $f_2$ 
is a periodic function.
In the traditional form of the PT model the ``friction force" is calculated as the average force that the driving has to apply in order to maintain a uniform driving velocity of $\dot\gamma$. This force is the one that stretches the spring, so the friction force $\sigma$ is calculated as
\begin{equation}
\sigma=\langle \dot\gamma t-e_2\rangle k
\label{sigma1}
\end{equation}
where the brackets notate a temporal average. In the yielding context, an equation like Eq. (\ref{pt1}) is taken as a mean field description of a spatially extended system. In the extended case, $\dot\gamma t$ represents the spatial (instantaneous) average of the value of $e_2$, namely
\begin{equation}
\dot\gamma t=\overline{e_2}
\label{sigma2}
\end{equation}
and the model equation must be written with the explicit inclusion of the externally applied stress $\sigma$ in the form
\begin{equation}
\dot e_2=f_2(e_2) +(\overline{e_2}-e_2)k+\sigma
\label{pt2}
\end{equation}
In any case, the difference between using Eqs. (\ref{pt1}) and (\ref{sigma1}), or (\ref{sigma2}) and (\ref{pt2}) is irrelevant as the difference amounts to a shift in the instantaneous value of $e_2$ by a constant, that does not alter the main feature of the flow curve (i.e., the $\dot\gamma$ vs $\sigma$ dependence) of the model.

We will extend now the PT model to include the possibility of volume fluctuations, described by a variable $e_1$. The value of $e_1$ will tend to evolve to its equilibrium value, with a dynamics mainly controlled by the value of the bulk modulus $B$ in the system. Considering an overdamped dynamics, this evolution must be of the form
\begin{equation}
\dot e_1=-Be_1
\label{e1}
\end{equation}
where the value $e_1=0$ was chosen as the equilibrium value in the system. Of course in this trivial form the evolution of $e_1$ is totally decoupled from the evolution of $e_2$, and there is no effect of $e_1$ (that adjusts at $e_1=0$ for all values of $\dot\gamma$) on $e_2$. 

\begin{figure}
\includegraphics[width=9cm,clip=true]{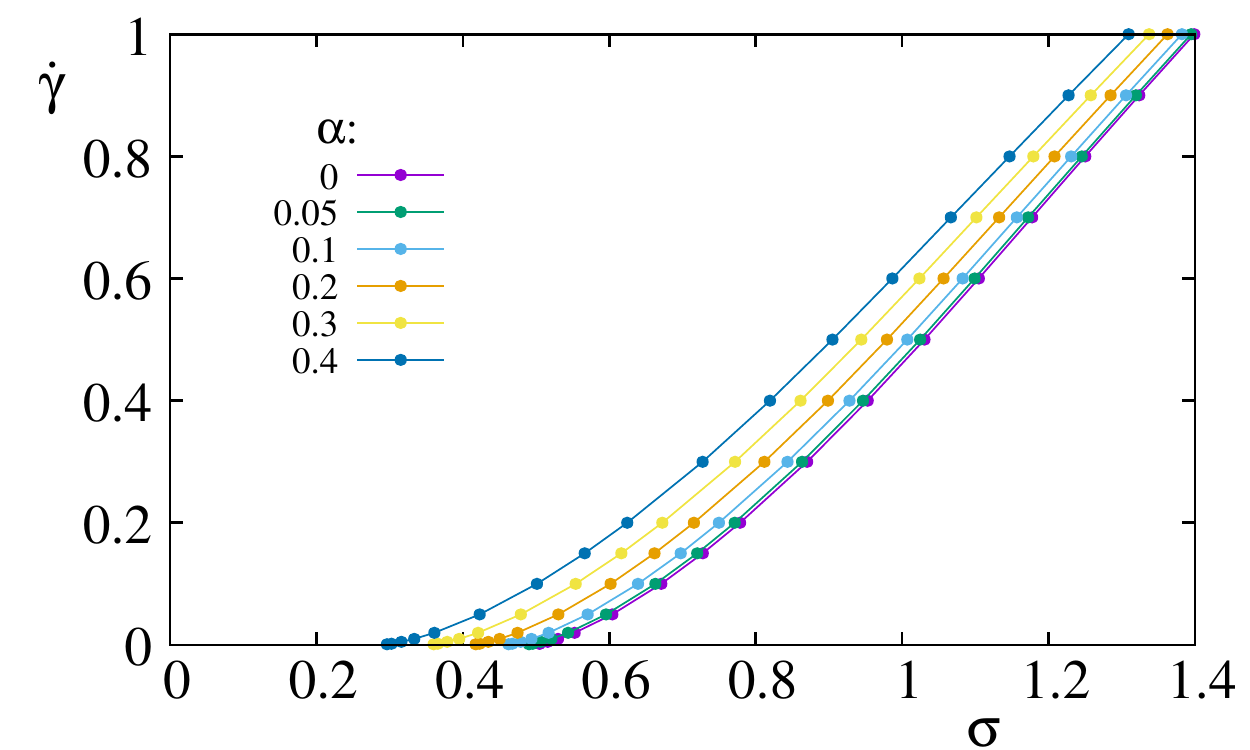}
\includegraphics[width=9cm,clip=true]{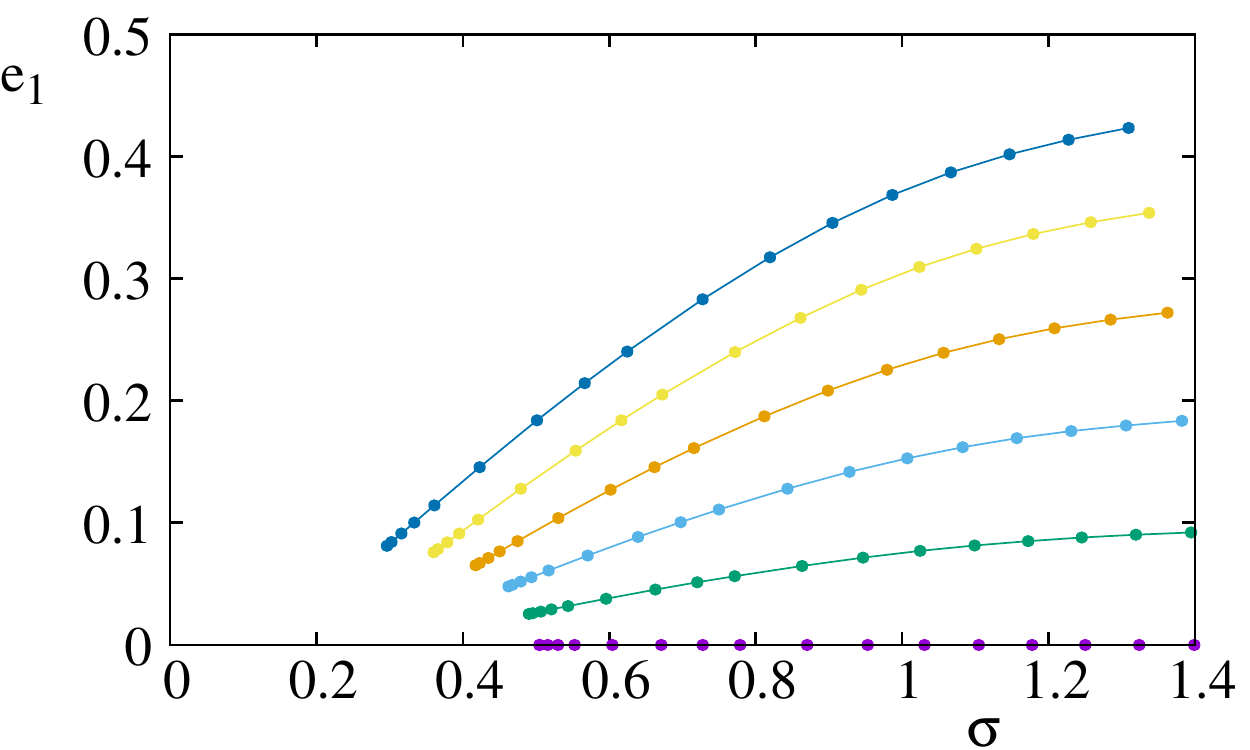}
\caption{(a) Flow curves of the one-particle model with coupling between shear and volume degrees of freedom. Different curves correspond to different values of $\alpha$ (Eq. \ref{v}) as indicated. (b) Corresponding curves of the temporal average of the variable $e_1$ representing the increase in system volume, caused by shear. The volume is larger for larger values of $\alpha$, and also increases with the value of strain rate $\dot\gamma$.}
\label{pt_dilatancy}
\end{figure}

Now we introduce a coupling between $e_1$ and $e_2$ that describes a possible dilatancy effect in the system. We expect that if $e_1$ increases, then the system yields more easily, therefore $e_1$ must have an effect on the force $f_2$. So we do the following. We consider a two dimensional potential energy $V(e_1,e_2)$ from which forces $f_{1,2}$ are obtained as the partial derivatives
$f_{1,2}=-\partial V/\partial e_{1,2}$. Therefore, the model equations will be 

\begin{eqnarray}
\dot e_1&=&-\partial V/\partial e_1\nonumber \\
\dot e_2&=&-\partial V/\partial e_2+ k(\dot\gamma t-e_2)
\label{modelo}
\end{eqnarray}
The form of $V(e_1,e_2)$ will dictate the form of the effective coupling between the modes $e_1$ and $e_2$.
We will use for $V$ the generic form 
\begin{equation}
V(e_1,e_2)= \frac B2e_1^2+U_1(e_1)U_2(e_2)
\end{equation}
Note first of all that we isolated the term $B e_1^2/2$ that corresponds to a sample with bulk modulus $B$. In the second term,  $U_2(e_2)$ will be an oscillatory potential with many minima along the $e_2$ axis.
In the case in which the factor $U_1(e_1)$ is a constant, this potential will provide the decoupled model in which $e_1$ and $e_2$ evolve separately.
However, if the value of $U_1(e_1)$ decreases as $e_1$ increases a dilatancy effect will be obtained. To lowest order we may consider that the change of $U_1$ is linear with $e_1$. We can conveniently model it by choosing

\begin{equation}
V(e_1,e_2)= \frac B2e_1^2+\left [1+\tanh(\alpha e_1)\right ] U_2(e_2)
\label{v}
\end{equation}
This form provides a corrugated potential on $e_2$ with an amplitude that tends to 0 when $e_1$ is positive and large, whereas saturating to an $e_1$-independent values for large negative $e_1$. Therefore $\alpha$ quantifies the coupling between $e_1$ and $e_2$. Coupling vanishes when $\alpha=0$.

In the next Section dealing with an extended system we will consider some stochastic form of the function $U_2$. Here, it suffices to consider a single periodic expression. In the past we have considered two particular forms of the $U_2$ term, described as the ``cuspy'' and ``smooth'' potential forms, defined as
\begin{eqnarray}
U_2(x)&=& (x-[x])^2/2~~~(\mbox {cuspy})\\
U_2(x)&=& \cos(x)~~~(\mbox {smooth})
\label{v_2}
\end{eqnarray}
(the square brackets in the first expression indicates the integer part function).
It is seen that the cuspy case consists of a concatenation of parabolic pieces.

Driving along the $e_2$ direction will produce the effect of modifying the value of $e_1$. 
We simulated Eqs. (\ref{modelo}), using expressions (\ref{v}), (\ref{v_2}) for the potential (for concreteness we work out the smooth potential case), at different levels of the coupling $\alpha$. The flow curve $\dot\gamma$ as a function of $\sigma$ and the temporarily averaged values of $e_1$ are reported in Fig. \ref{pt_dilatancy}.
We see how the volume $e_1$ increases as $\dot\gamma$ (or $\sigma$) increases. This is the dilatancy effect. 
The effect is stronger as the value of $\alpha$ increases. Note also how the flow curve for larger values of $\alpha$ shifts to the left (and in particular the critical stress reduces) at larger $\alpha$ when $e_1$ is positive. 

The dilatancy effect in the PT model is the germ of a couple of interesting features that we will encounter in the spatially extended model, and that are analyzed in the next two sections.

\section{Flow curves and orientation of shear bands in the presence of dilatancy}

In a spatially extended system, and in the absence of volume-shear coupling, correlated slips occur at 45$^\circ$ with respect to the principal axis of the stress tensor, as this is the orientation in which the maximum shear stress occurs. 
However, in the presence of volume-shear coupling, this angle changes.\cite{kamran,procaccia}
We address this effect now.

\begin{figure}
\includegraphics[width=9cm,clip=true]{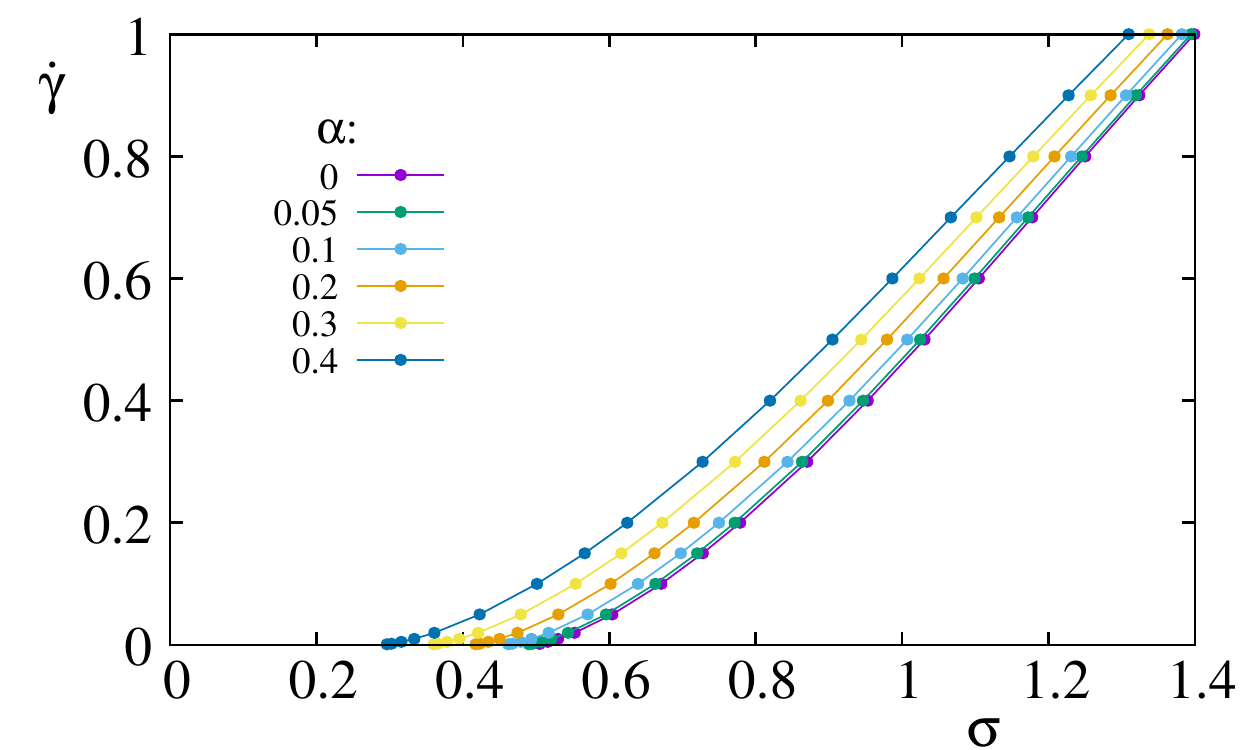}
\includegraphics[width=9cm,clip=true]{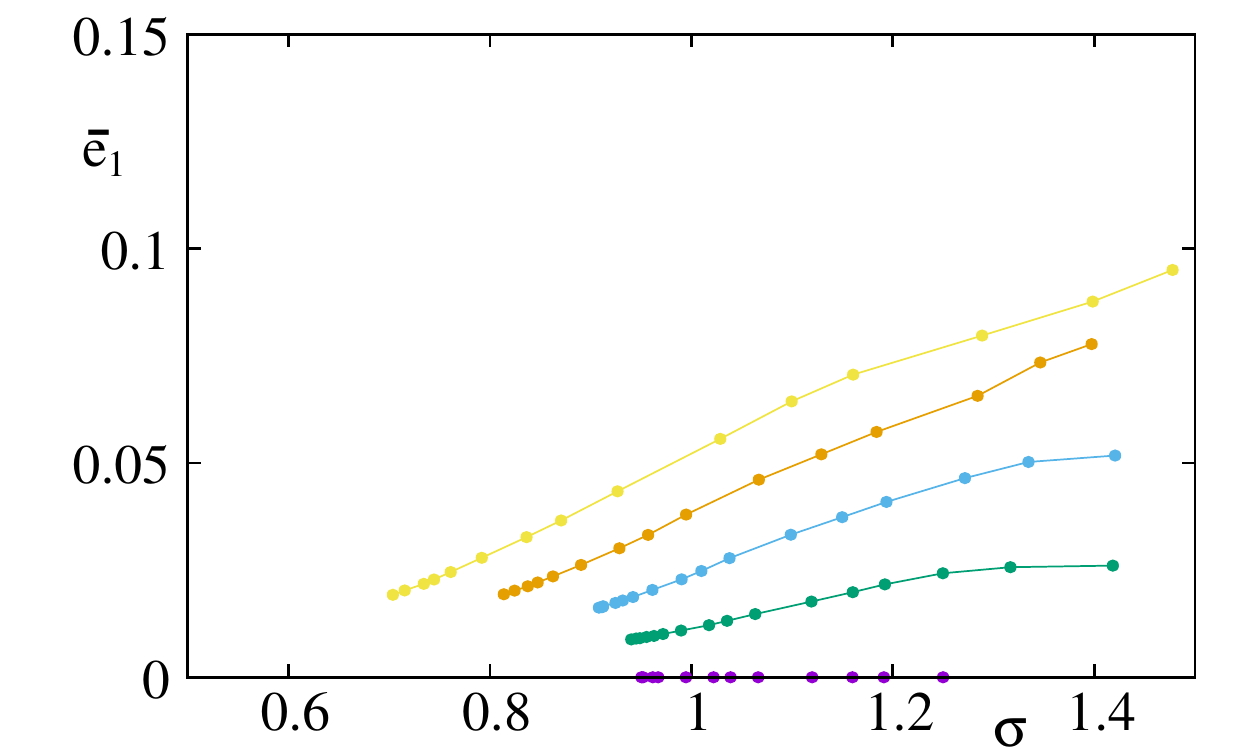}
\caption{(a) Flow-curves ($\dot\gamma$ vs $\sigma$) at different values of the volume-shear coupling parameter $\alpha$, as indicated. (b) The corresponding curves for the average volume change. The qualitative behavior of the model is consistent with the results obtained in the one-particle model (Fig. \ref{pt_dilatancy}).}
\label{flow_curves}
\end{figure}

The one-site modeling of the previous section is the base for a spatially extended model in 
which each site is characterized by its strain and its volume degree of freedom. Details of 
the model have been presented elsewhere\cite{jagla_2007,jagla1,jagla2}, and here we only state its more relevant characteristics. We work with three 
functions $e_1({\bf r})$, $e_2({\bf r})$, $e_3({\bf r})$, that represent one volume and two 
shear degrees of freedom at each position. Volume-shear coupling is obtained by the use of a 
potential energy qualitatively similar to that in Eq. (\ref{v}), but also with some differences 
due in part to technical reasons.
First of all, taking into account that physically $e_2$ and $e_3$ differ only 
by a spatial rotation of $45^\circ$, 
we should consider $(e_2^2+e_3^2)$ as the variable to couple to $e_1$ through a parameter $\alpha$ 
similar to that used in the previous section. However notice that in our implementation using finite 
differences onto a square lattice, the variables $e_1$ and $e_2$ are calculated on the nodes of this 
lattice, whereas $e_3$ is defined at the centers of the plaquettes. This makes a possible coupling 
between $e_1$ and $e_3$ a non-local one in the present numerical scheme, producing spurious effects that require special care to 
be avoided. This is certainly a point that will deserve further attention. However, since the 
external driving is applied on the $e_2$ direction we may also expect that the main effect we 
are interested in can be captured by coupling $e_1$ to $e_2$ only, leaving $e_3$ evolve without any 
coupling to $e_1$.
Therefore we will model the local potential energy as
\begin{equation}
V(e_1,e_2,e_3)= \frac B2e_1^2+\left [1+\tanh(\alpha e_1)\right ]U_2(e_2)+U_3(e_3)
\label{v3}
\end{equation}
The functions $U_2$, $U_3$ will be taken to be of the ``smooth'' type, but with a stochastic ingredient, this stochasticity being totally uncorrelated in different spatial positions. We proceed as follows. We define a value 
$e_0$ (either $e_{20}$ or $e_{30}$ for $e_2$ or $e_3$) and define the corresponding $U$ to be given by
\begin{equation}
U= 1-\cos\left (   \frac{(e-e_0)\pi}{\Delta}\right )
\end{equation}
in the interval $|e-e_0|<\Delta$,
where $\Delta$ is a random variable taken from a uniform distribution between 0.5 and 1.
We monitor along the simulation the values of $e-e_0$ and compare against the value of $\Delta$. Each time we detect
$\pm(e-e_0)>\Delta$ a new value $\Delta_{new}$ is defined, and $e_0$ is redefined to $e_0\pm (\Delta+ \Delta_{new})$. In this form we implement an independent stochastic disordered potentials for $e_2$ and $e_3$ at every position in the sample. The equations of motion of the model are taken to be of the overdamped form
\begin{equation}
\dot e_i=-\frac {dV}{de_i}
\end{equation}
that however must be solved taking into account that there is a constrain among $e_1$, $e_2$, and $e_2$ that must be fulfilled (the so called St Venant constrain \cite{stvenant}), and that is imposed by the appropriate use of Lagrange multipliers \cite{jagla_2007,jagla1,jagla2}. This constrain is the ingredient that generates the well known long-range elastic Eshelby interaction in the system.

\begin{figure}
\includegraphics[width=6cm,clip=true]{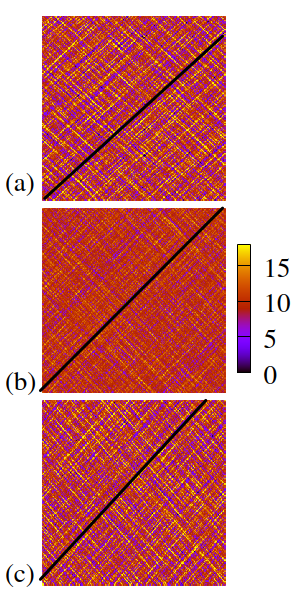}
\caption{Snapshots of accumulated deformation in the system in a fixed strain window at different values of the volume-shear coupling parameter: $\alpha=0.2$ (a), 0 (b), and -0.2 (c). Note the change in the orientation of the slip lines (highlighted by the diagonal or off-diagonal straight black lines) as a function of $\alpha$.}
\label{angulos}
\end{figure}

\begin{figure}
\includegraphics[width=9cm,clip=true]{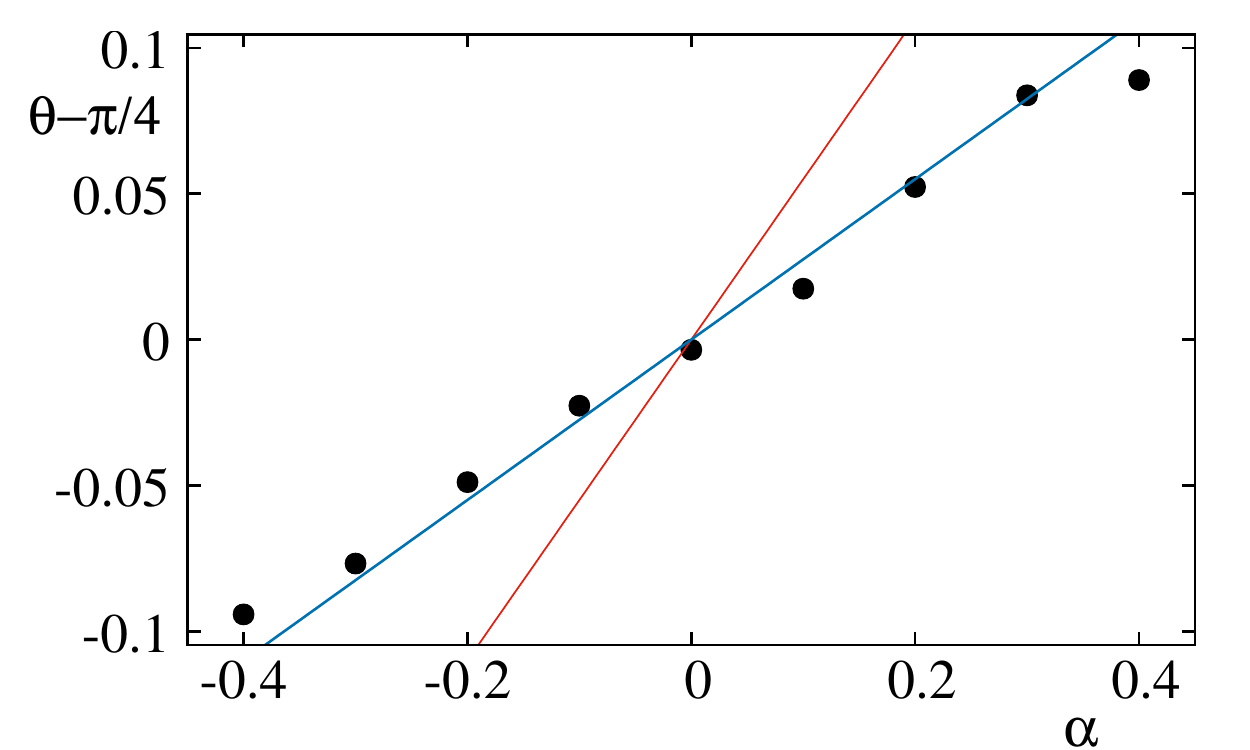}
\caption{Angle of maximum correlation between slips in the sample as a function of $\alpha$. Dots are the measured values. Red line is the result of the Mohr Coulomb theory contained in Eq. (\ref{factor2}). Blue line is the result of the theory in \cite{kamran}.
}
\label{angulos2}
\end{figure}

We present some flow and volume curves for this model in Fig. \ref{flow_curves}. We observe in the extended system the same qualitative features obtained for the one particle case (Fig. \ref{pt_dilatancy}), namely a shift of curves to the left and an increase in system volume as $\alpha$ is increased. 
A more detailed inspection of the spatial distribution of deformation reveals a particular effect associated to a non-zero value of 
$\alpha$. 
In Fig. \ref{angulos} we see plots of accumulated strain $e_2$ in the system, at a relatively large value of $\dot\gamma=0.1$, in systems with different values of $\alpha$. 
In Fig. \ref{angulos} 
the angle of correlation of plastic slips is clearly visible,
and is observed that this angle deviates from $45^\circ$ when $\alpha\ne 0$. By calculating a correlation function on spatial configurations as those in Fig. \ref{angulos}, the angle of maximum correlation can be determined with good precision.
The result as a function of $\alpha$ is seen in Fig. \ref{angulos2}.
The angle is $45^\circ$ in the case $\alpha=0$, but deviates from this value when $\alpha\ne 0$, the deviation being linear with $\alpha$.

The orientation correlation of plastic activity (and ultimately the orientations of shear bands) with respect to principal stress directions is an important feature in many practical situations, so the reason for this reorientation is worth to be explored in more detail.
When $\alpha=0$ the failure planes are determined by the condition of maximum shear stress, and this always occur at 45$^\circ$ with respect to the principal axis (in the present simulations they being the $x$ and $y$ axis, we will take $\sigma_x>\sigma_y$).
However, if $\alpha\ne 0$ the stress normal to the failure plane plays a role as a reduction in normal stress favors slip by reducing the critical stress. Therefore there is rotation  of the slip planes with a normal that tends to align with the axis in which the compressive stress is minimum. 
The standard way to quantitatively calculate this rotation is the Mohr-Coulomb approach\cite{nedderman}, that goes as follows. Considering potential slip planes appearing at an angle $\theta$ with respect to the $x$ axis, the normal 
($\sigma_n$) and shear ($\tau$) stresses on these slip planes are given by

\begin{eqnarray}
\sigma_n(\theta)&=&p-\sigma\cos(2\theta)\label{st1}\\
\tau(\theta)&=& \sigma \sin(2\theta)\label{sigma-tau}
\label{st2}
\end{eqnarray}
where $p\equiv(\sigma_x+\sigma_y)/2$, and $\sigma\equiv(\sigma_x-\sigma_y)/2$.
The Mohr-Coulomb theory considers that a sample is in static equilibrium as long as the frictional equilibrium condition
\begin{equation}
\tau(\theta)<\sigma_n(\theta) \mu +c
\end{equation}
is satisfied for all $\theta$, with $\mu$ and $c$ being sample parameters named the internal friction coefficient and cohesion, respectively. 
Inserting Eqs. (\ref{st1},\ref{st2}) the stability condition can be written as
\begin{equation}
(\sin(2\theta)+\mu\cos(2\theta))\sigma<c+p\mu
\label{sigmac}
\end{equation}

The Mohr-Coulomb criterion states that actual slips will occur at the angle $\theta$ for which
this condition first becomes an equality. It is clear that this will occur at a value of $\theta$ for which 
$(\sin(2\theta)+\mu\cos(2\theta))$ is a minimum, predicting a slip angle given by
\begin{equation}
\theta=\frac 12 \atan (\mu^{-1})
\label{tetamu}
\end{equation}
Introducing the internal friction angle of the sample $\phi$ such that $\mu=\tan \phi$, $\theta$ can be written as
\begin{equation}
\theta=\pm \left( \frac \pi 4 -\frac \phi 2\right)
\label{theta02}
\end{equation}

In order  to calculate the value of $\phi$ for a given sample, the {\em failure envelope} of the sample
must be determined. By replacing back Eq. (\ref{tetamu}) into (\ref{sigmac}) it is obtained that failure (or yielding) occurs when
\begin{equation}
\sigma>\frac {c+p\mu}{\sqrt{1+\mu^2}}=c\cos(\phi)+p\sin(\phi)
\end{equation}
Therefore, a determination of the stability region in the $p,\sigma$ plane allows to determine the values of $\phi$ and $c$, and calculate the slip angle $\theta$.

\begin{figure}
\includegraphics[width=9cm,clip=true]{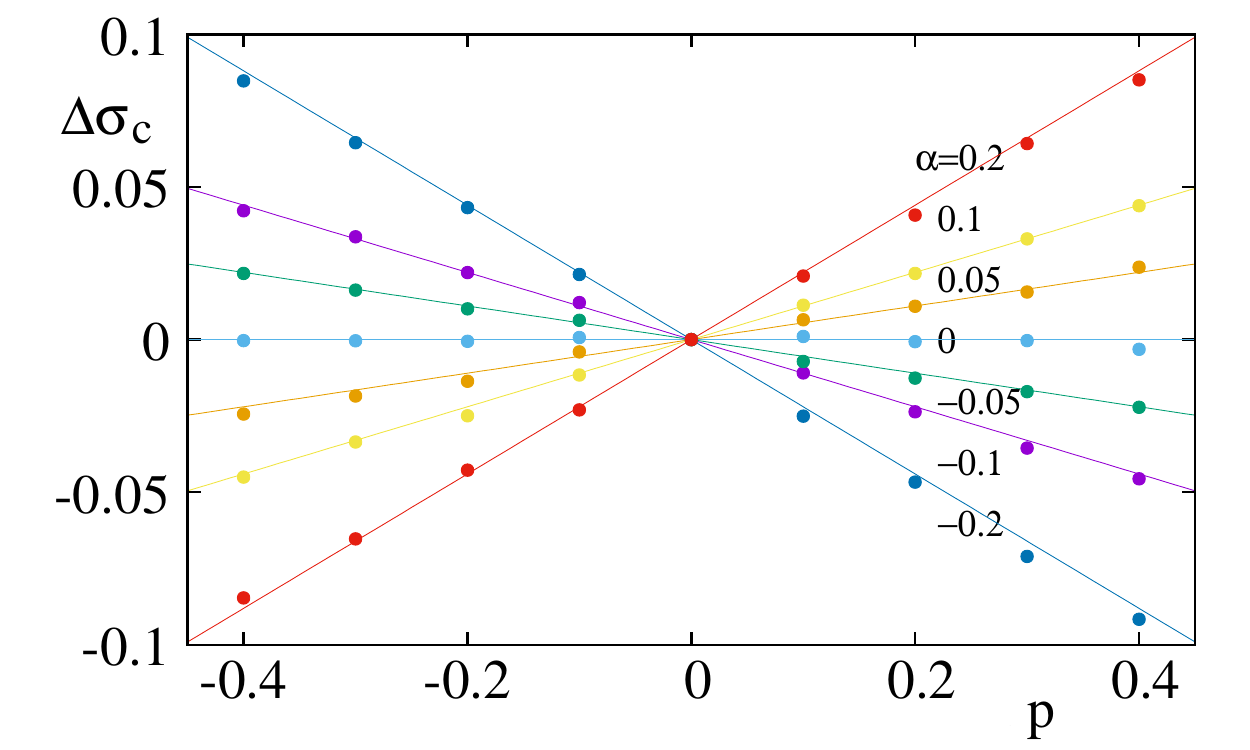}
\caption{Variation of the critical force $\Delta\sigma_c\equiv (\sigma_c(p)-\sigma_c(0))$ as a function of the compressive stress $p$. Dots are the results of the simulations for different values of $\alpha$. Lines are plots of the function $1.1 \alpha p$, which provides a very good fitting of the numerical results.}
\label{f4ymedio}
\end{figure}

A compressive stress $p$ can be easily added in our numerical model, and the critical value of $\sigma$, namely $\sigma_c$ calculated as a function of $p$ for different values $\alpha$. The results (shown in Fig. \ref{f4ymedio}) 
are compatible with a linear dependence of $\phi$ on $\alpha$ given by

\begin{equation}
\phi\simeq 1.1\alpha
\label{fialfa}
\end{equation}
This expression can now be used in Eq. (\ref{theta02}) to obtain
\begin{equation}
\theta \simeq\pm\left(\frac \pi 4-0.55 \alpha\right)
\label{factor2}
\end{equation}
which is the Mohr-Coulomb prediction for the angle of plastic slips in terms of the volume-shear coupling parameter $\alpha$. This linear function is plotted in red on top of the numerical results in Fig. \ref{angulos2}. It reproduces the linear trend of the numerical results, but it overestimates the numerical values by about a factor of two. This kind of behavior goes in the same direction as experimental results, where it is observed also that Mohr-Coulomb theory overestimates the angle of shear bands with respect to $\pi/4$ \cite{bardet}.

Recently, Karimi and Barrat \cite{kamran} have presented a detailed analysis of the orientation of shear bands, concluding also that the macroscopic Mohr-Coulomb criterion is not accurate in quantitative predicting the orientation of slip. Instead, they rely on an analysis in term the elastic coupling between plastic distortions and the {\em local} internal friction angle $\phi_0$ to conclude that the orientation of the slip planes $\theta$ is given by
\begin{equation}
\cos(2\theta)=\frac 1 2 \sin(\phi_0)
\end{equation}
or approximately 
\begin{equation}
\theta \simeq \pm\left( \frac \pi 4- \frac {\phi_0} 4\right)
\label{fis4}
\end{equation}
in cases in which $\phi_0$ is not too large\cite{footnote}.
The local internal friction angle ${\phi_0}$ plays the same role as $\phi$ but for an elementary piece of the material, and does not include effects of the elastic interaction that are effectively incorporated in the macroscopic $\phi$.
In our samples the value of $\phi_0$ is directly related to the shear volume coupling parameter $\alpha$. We have measured this dependence by numerically finding the failure envelope now for a single element in the system, namely we ran the one-site model of Section II using the stochastic potential of the full model, and calculating the critical stress as a function of the normal stress. The slope of this dependence provides the value of $\sin(\phi_0)$. We have found that $\phi_0$ can be adjusted as
\begin{equation}
\phi_0\simeq 1.1\alpha
\end{equation} 
This means that in our model, and within the numerical precision, the value of $\phi_0$ coincides with the macroscopic value $\phi$ (Eq. (\ref{fialfa})). Yet, the theory in \cite{kamran} suggests to use (\ref{fis4}) to calculate $\theta$, and this gives a result that has half the slope with respect to $\alpha$ that the Mohr-Coulomb theory, finding now results (Fig. \ref{angulos2}, blue line) the adjust reasonably well the experimental correlation observed for plastic slips.

\section{Non-uniform yielding caused by volume-shear coupling}

\begin{figure}[h]
\includegraphics[width=8cm,clip=true]{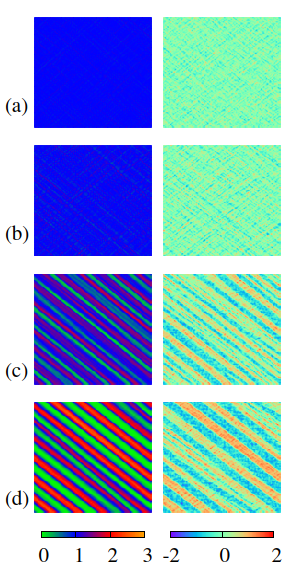}
\caption{Accumulated strain deformation (left column, the scale value 1 corresponds to the nominal average deformation in all cases) and instantaneous values of $e_1$ (right column) in simulations at the same value of $\dot\gamma=0.1$, in systems with different values of shear-volume coupling parameter $\alpha$, namely $\alpha=0.1$ (a), $0.2$ (b), $0.3$ (c), and $0.4$ (d) (system size is 128 $\times$ 128). The long term deformation is uniform in the first two cases, but is inhomogeneous in the last two.}
\label{separacion}
\end{figure}

An important phenomenon that may occur in the flowing of amorphous materials is localization of the deformation. In some cases (as it occurs in metallic glasses, see next section) this localization is a non-equilibrium phenomenon typically associated to the preparation of the sample. In other cases, this localization occurs at equilibrium and this is the situation that is addressed here.  In Ref. \cite{lequeux} a stability analysis was performed for a flowing system that has, apart from the normal shear variables, additional scalar degrees of freedom taken to represent 
the concentration of solute molecules in the material. 
The analysis made is quite general, and essentially applies also to our case in which the additional scalar variables can be considered to be the value of the local density in the system. 
Restricted to a two-dimensional situation appropriate to our case, the result of the analysis indicates that there are two different kinds on instabilities. In one of them, instability is purely mechanical, with the density degree of freedom playing no important role. The signature of such an instability is a flow curve that has a reentrance, namely a region in which stress diminishes as strain rate increases. This kind of instability is analogous to that producing phase separation in a one component fluid, such as in the liquid-gas transition. In physical terms, this instability occurs in the following way. If a layer of the system that is shearing in a stationary way suffers a perturbation that increases its strain rate, then it would require less stress to maintain the new strain rate. But as the applied stress is kept fixed, this will produce an additional increase of strain rate, leading to an instability.  
This situation is known to occur for instance in cases in which there are aging mechanisms in the system\cite{jagla_2007,picard,olmsted,divoux,coussot_prl_2002,mujumdar,martens}. It may thus happen that non-flowing regions keep a well aged state that is more rigid, and therefore do not flow, while flowing regions, being unable to age sufficiently, have a lower critical stress and are maintained in a flowing state. 

In addition to this kind of mechanical instability, it was found \cite{lequeux} that the coupling to the concentration degree of freedom can produce a breaking of the uniform flow, even in the absence of any strong feature of the flow curve. This instability is qualitatively understood in the following way.\cite{besseling,fall,ovarlez} 
If on an originally uniform yielding state, a fluctuation in some part of the sample produces an increase in strain rate, then that part of the system will also experience an expansion 
which in turn produces a decrease in the effective viscosity and a further increase in the strain rate. This process can eventually lead to an instability which however does not manifest clearly in the form of the flow curve of the system. 

We have observed this kind of instability in the present model, in the case in which $\alpha$ is large enough. 
The numerical evidence is contained in Fig. \ref{separacion}. There in the left column we observe plots of the accumulated deformation in the system in a rather long time window
at a constant value of $\dot\gamma=0.1$, and for different values of $\alpha$. For low $\alpha$ ($\alpha\lesssim 0.2$) the deformation is uniform, whereas for larger $\alpha$ we clearly see that there are regions in the system that have not yielded at all, showing the non-uniform yielding in this case. The instantaneous values of the volume variable $e_1$ across the system (right column in Fig. \ref{separacion}) show also a clear correlation with the deformation in the large $\alpha$ cases, with the sample flowing in regions in which the local volume is well above the average value. Stuck regions instead correspond to compressed sample regions, with values of $e_1$ below the average.  The flow curves corresponding to these cases are precisely those shown in Fig. \ref{flow_curves}, and in fact, we do not see there any sign of the spatial localization that is observed in the snapshots of Fig. \ref{separacion} at large $\alpha$.

\section{Volume-shear coupling in annealed samples. The case of metallic glasses}

The effects of volume-shear coupling that have been discussed so far all correspond to situations of dynamical equilibrium in the system, or in other words, they all describe stationary situations.
There are other situation however, 
in which we are interested in transient (although probably long-lived) properties of the system. In many cases this occurs when we do an experiment on a sample prepared by some ad-hoc protocol, that leaves it in an out of equilibrium configuration. In concrete we want to refer to the case of metallic glasses.\cite{tang,greer,ogata,zhong}
They  are usually prepared by annealing from a melt, and its configuration cannot by any means be considered as an equilibrium one. When the metallic glass is submitted to an external load, the appearance of a shear band is a highly non-equilibrium process. Yet the kind of models we have presented can be used to model this scenario, and in particular the effect of volume fluctuations on the behavior of this kind of materials.

The key  to mimic the behavior of a metallic glass under shear is to start with an initial configuration that appropriately incorporates the annealed nature of the experimental initial state. In our model we can produce a starting configuration that is more stable than the stationary one after long deformation by adjusting the local disorder potentials at different sites\cite{jagla_2007}. The local potentials are characterized by the values of $e_{20}$ and $e_{30}$, defining the local equilibrium strain,  and the corresponding values of $\Delta$ measuring the extent of the corresponding basin. In stationary situations as those considered so far, the values of $e_0$ have some dynamically generated dispersion, and $\Delta$ is randomly chose from a uniform distribution between 0.5 and 1.0. We can construct a more stable starting sample by appropriately modifying $e_0$ and/or $\Delta$.

The possibility that we will implement here  is to anneal the sample (before strain is applied) by allowing the values of 
$e_{20}$ and $e_{30}$ to relax towards more stable configurations. This is the scheme that was originally used in Ref. \cite{jagla_2007}. Here we apply the same protocol in the presence of volume-shear coupling.
In concrete, we first generate a stationary state by running a long simulation with a small value of $\dot\gamma=0.001$. In this situation we stop driving (setting $\dot\gamma=0$) and perform an annealing process that consists in adapting dynamically the values of $e_{20}$ and $e_{30}$ according to
\begin{eqnarray}
\dot e_{20}\sim \nabla^2 e_{20}\\
\dot e_{30}\sim \nabla^2 e_{30}
\end{eqnarray}
while at the same time continuing to relax the values of $e_1$, $e_2$, and $e_3$ following the standard evolution equations of the model. This produces, after annealing for some period of time, a configuration that is more relaxed than the stationary state, because now the values of $e_{20}$ and $e_{30}$ are strongly correlated spatially, and the surface is effectively more pinned.
Finally,  we stop the relaxation of $e_{20}$ and $e_{30}$, generating the starting configuration. 
We present results in the extreme case in which relaxation is performed during a very long time, and therefore, the values of $e_{20}$ and $e_{30}$ acquire a uniform value across the system. 
From this initial configuration strain is increased at a low strain rate $\dot\gamma=0.001$, and the stress is numerically evaluated.

\begin{figure}
\includegraphics[width=8cm,clip=true]{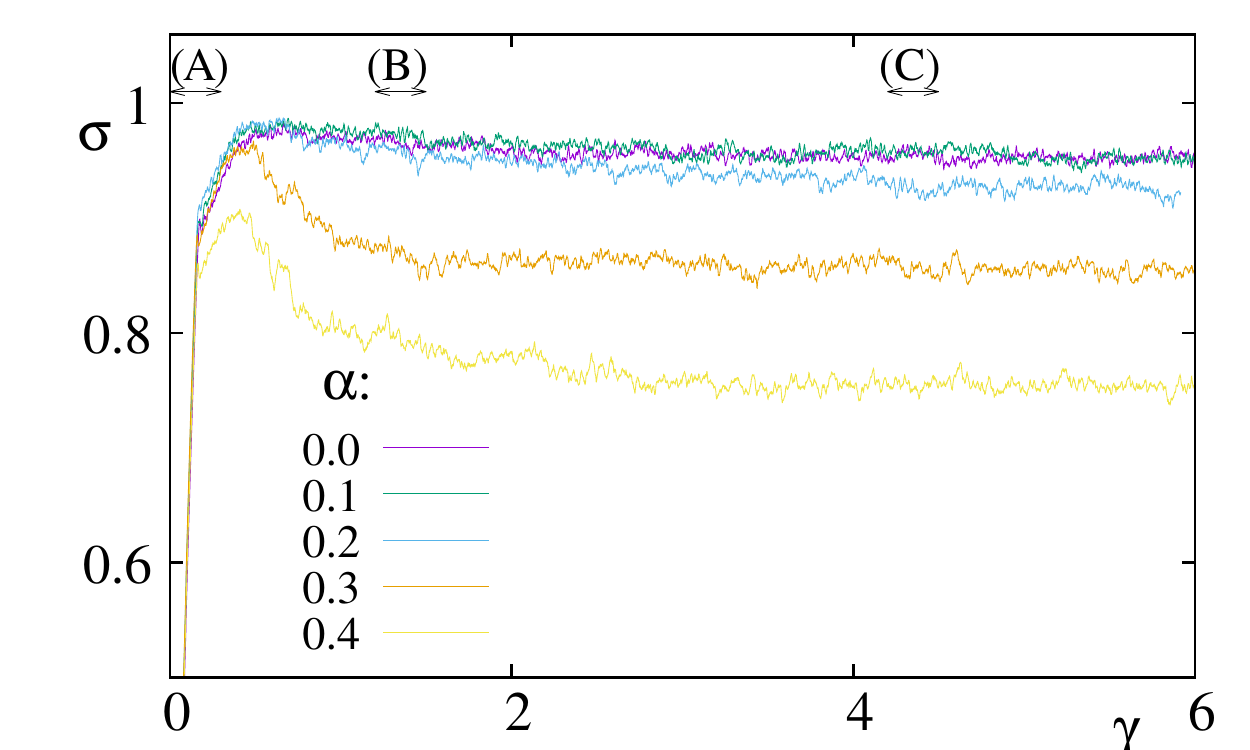}
\caption{Stress-strain curves (using $\dot\gamma=0.001$) in samples with different values of $\alpha$ as indicated, after an annealing in the initial state, as described in the text. All curves display a stress overshoot above the asymptotic values, but the effect is much stronger in samples with large $\alpha$.}
\label{f8}
\end{figure}

\begin{figure}
\includegraphics[width=9cm,clip=true]{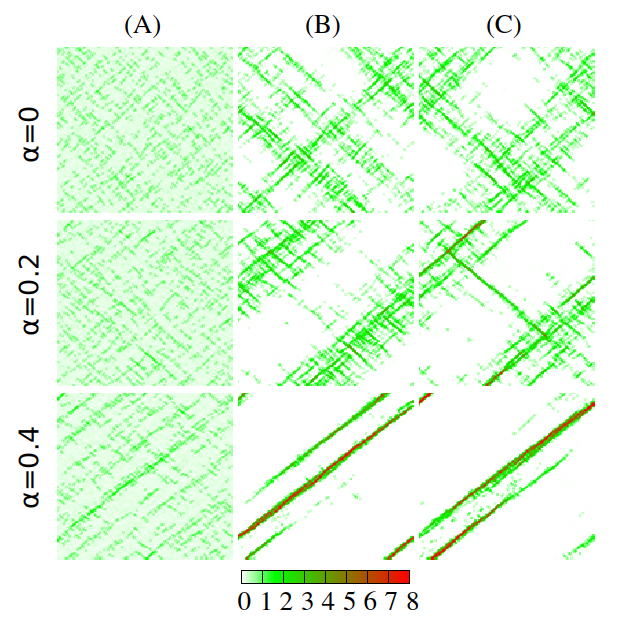}
\caption{Accumulated spatial distribution of strain in the three strain windows indicated in Fig. \ref{f8}, for three different values of $\alpha$. Although for all values of $\alpha$ we observe strain localization, the effect for larger $\alpha$ produces a much narrow shear band.}
\label{f9}
\end{figure}

The results obtained are shown in Fig. \ref{f8} for systems with different intensity of volume-shear coupling. All curves in Fig. \ref{f8} display a stress overshoot that is the hallmark of annealed samples that are more stable than the asymptotic dynamic equilibrium state. Yet we see that the amplitude of the stress overshoot is much more pronounced in samples with larger value of $\alpha$. In order to have a deeper understanding of the effect of $\alpha$ we must investigate the local distribution of deformation in the samples. 
In Fig. \ref{f8} we have indicated three strain windows, labelled (A), (B), and (C).
The local distribution of accumulated strain in each of these windows is indicated in Fig. \ref{f9} for three different values of $\alpha$, namely $\alpha=0, 0.2, 0.4$. 

\begin{figure}
\includegraphics[width=5cm,clip=true]{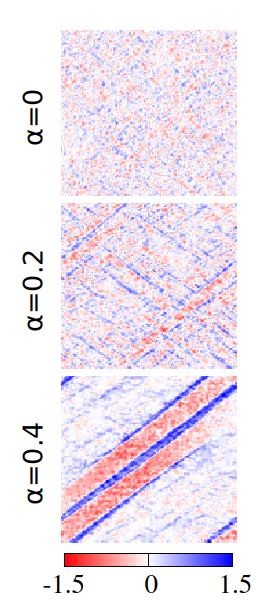}
\caption{Instantaneous distribution of the volumetric $e_1$ variable, right after the time window (C), for the three different values of $\alpha$ presented in Fig. \ref{f9}. The position of the shear band in the system can be inferred from the distribution of $e_1$ in the case in which $\alpha$ is large.}
\label{f10}
\end{figure}

The case $\alpha=0$ serves as a reference, and is similar to that studied in Ref. \cite{jagla_2007}. We emphasize that even in this case, an overshoot is observed in the stress-strain curve (Fig. \ref{f8}). In the strain window (A), when stress is still increasing in the system, we observe rather isolated plastic events that do not yet organize across the sample. In strain windows (B) and (C) instead, when the stress peak has been overpassed, the plastic strain is localized in a spatial region that can be termed a shear band. The reason for this behavior is that in the shear band the system was forced to escape from the initial annealed state, and its critical stress is now lower than in the still annealed part, that is non flowing. This stabilizes the shear band in the system.  
This shear band widens as $\sim \gamma ^{1/2}$ when strain is increased \cite{jagla_jstat_2010,falk1,falk2}. Notice that the strain  localization at $\alpha=0$ is not observable in the distribution of the volume variable $e_1$ (Fig. \ref{f10}(a)).

The case of a finite shear-volume coupling ($\alpha>0$) produces some quantitative differences and also some qualitative
new phenomena. First of all, in Fig. \ref{f8} we clearly see how the increasing part of the stress-strain curve is not much dependent on the value of $\alpha$, and in fact, the spatial distribution of the deformation in strain window (A) shows a rather uniform deformation distribution independently of $\alpha$. However, after the stress peak is overcome, the stress decrease is much larger in samples with large $\alpha$. 
The analysis of the spatial distribution of deformation reveals that 
the accumulated strain at finite $\alpha$ shows  a stronger tendency to become more localized, and a much weaker tendency to widen as strain increases. Most remarkably, for values of $\alpha$ where we had previously observed an instability of the homogeneous deformation situation ($\alpha \gtrsim 0.2$, see Fig. \ref{separacion}), we now observe that only a very thin shear band is activated in the system, with all the rest remaining in a rigid configuration. This is very clear in the plots corresponding to $\alpha=0.4$ in Fig. \ref{f9}. Notice that as the shear band does not run exactly along the diagonal (see Section III), and because of the periodic boundary conditions, there are some ``mirror images" of the shear band, giving the impression of a few different shear bands in the system.

We think the most remarkable effect of a finite $\alpha$ manifests on the spatial distribution of the $e_1$ variable (Fig. \ref{f10}). While for $\alpha=0$ we did not have any clue of the strain localization from the distribution of $e_1$, in the finite $\alpha$ case, particularly when $\alpha \gtrsim 0.2$ this is not so. In fact, in Fig. \ref{f10}(c) we clearly observe that the localization of the shear band corresponds to a region where the system is expanded (therefore facilitating the shearing of the system in this region). We emphasize that if we stop increasing the strain in the system, or even if strain is reduced as necessary to reach a state of zero stress, the spatial distribution of $e_1$ remains essentially at the metastable configuration shown in Fig. \ref{f10}. Therefore, the localization of the shear band where the system has yielded can be evaluated after the full process took place, by localizing the regions of the sample that are expanded with respect to the starting configuration.  
 

\section{Summary and conclusions}

In this work, we have introduced a mesoscopic model for the yielding behavior of amorphous materials incorporating the possibility of volume-shear coupling. This ingredient is acknowledged to have important experimental consequences, but
has only rarely \cite{kamran} been previously considered in numerical models of the yielding transition. Volume-shear coupling produces a number of effect in the phenomenology of yield stress materials. It generates a change in the direction of the shear bands in the system. Our numerical simulations clearly show this angular deviation. We have applied the classical Mohr-Coulomb approach and the theory in Ref. \cite{kamran} to predict the deviation, and find
that the prediction in \cite{kamran} nicely fits our numerical results, whereas the Mohr-Coulomb theory tends to overestimate the reorientatin effect by about a factor of two.
Also, we have observed that when the volume-shear coupling is strong enough, there is a tendency of the system to generate a non-uniform deformation. IN this case, there are regions that become more expanded, and therefore yield more easily, while others remain in a more compact configuration, and do not yield at all. Finally, we have briefly discussed the
effect of volume-shear coupling in cases in which the sample is initially annealed to obtain a starting configuration that is more stable than the asymptotic configuration that appears as a stationary state after a very long deformation
time. This situation is particularly adapted to the phenomenology of metallic glasses. 
Now, when the sample is strained, the stress in the system develops a peak before smoothly decaying to the asymptotic stress value. The case of no volume shear-coupling has been studied in previous works \cite{jagla_2007,falk1,vandembroucq} and it was observed that after the strain corresponding to the stress maximum is exceeded, the deformation in the sample localizes in a shear band. This band progressively widens with further deformation as discussed in Ref. \cite{jagla_jstat_2010,falk2}.
Now in the case in which there is volume-shear coupling, particularly when this is large, we observe that the shear band that appears in the system is very thin, and we do not detect clearly that it widens as deformation proceeds. We observed that in the region of this thin band the system is expanded, and this expansion remains even if deformation is stopped and stress relaxes back to zero. 
All these findings point to the utility of the present kind of model in the study of shear stress materials for which shear-volume coupling is important.

\section{Acknowledgments}
I thank Ezequiel Ferrero for helpful stimulating discussions, and Kamran Karimi for bringing to my attention the theory in Ref. \cite{kamran}.


\begin{thebibliography}{1}
\bibitem{footnote}Note that our variables $\phi$ and $\phi_0$ are notated in Ref. \cite{kamran}
as $\phi_r$ and $\phi$, respectively. Also in Ref. \cite{kamran}, an effect of ``dilatancy" on the relation between $\theta$ and the internal friction angle is predicted. Although our shear volume coupling mechanism induces dilatancy of the sample we think the effect discussed in \cite{kamran} does not apply to the results in this Section, because here the sample is dilated permanently and uniformly.
\end{thebibliography}
\end{document}